\documentclass[12pt]{article}
\usepackage{epsfig}
\usepackage{subfigure}
\usepackage{amsmath}
\usepackage{amssymb}

\begin{document}
\title{Hints of Energy Dependences in AGASA Extremely High Energy
Cosmic Ray Arrival Directions}
\author{William~S.~Burgett and Mark~R.~O'Malley\\
Department of Physics\\ University of Texas at Dallas\\
Richardson,TX  75083\thanks{email: burgett@utdallas.edu}}
\date{}
\maketitle
\abstract
\noindent
A correlation and probability analysis of the distribution 
of arrival directions for a  sample of
AGASA events reported to have energies above $4\times10^{19}$~eV shows the 
small scale clustering to remain significant at the $99.5 - 99.9\% \,$CL and to be consistent 
with previous results. For the sample taken as a whole, there are no departures
from either homogeneity or isotropy on angular scales 
greater than 5 degrees. The sample of events with $E \geq 6\times10^{19}\,$ 
eV contains no pairs with separations $< 4$ degrees. 
Cross correlating subsamples partitioned by energy reveals three uncorrelated
distributions in the intervals $4 - 5 \times10^{19}\,$eV, $5 - 8 \times10^{19}\,$eV,
and greater than $8 \times 10^{19}\,$eV. The partition with 
$5 \leq E < 8\times10^{19}\,$eV is correlated with the supergalactic equatorial plane while
the other two groups are statistically consistent with isotropic distributions.
The presence of three distinct energy-partitioned groups of events could reflect
possible changes in primary 
composition, different source distributions, differing levels of GZK losses,
or deflection effects of magnetic fields.

\setlength{\parskip}{2.0ex plus0.5ex minus0.5ex}
\noindent
PACS number(s): 95.85.Ry, 98.70.Sa 

\vspace{18mm}
\centerline{Accepted for publication in Phys. Rev. D}

\newpage
\maketitle
\section{Introduction}
\noindent
Investigating cosmic ray airshowers induced by extremely high energy cosmic rays (EHECRs, 
$E \: \gtrsim \: 4\times10^{19}$ eV) remains an area of
intense interest because of an
apparent paradox posed by the onset of the Greisen-Zatsepin-Kuzmin (GZK) effect at these 
energies. The GZK effect provides an efficient energy loss mechanism 
for sufficiently energetic cosmic ray protons interacting 
with cosmic microwave background~(CMB) photons~\cite{greisen,zatsepin}. 
Specifically, for cosmic ray primaries of energy $E \: \gtrsim \: 4\times10^{19}$~eV,
photopion production is greatly enhanced through the excitation of the $\Delta^{+}$(1232 MeV)
resonance. As the mean free path for this process is $\sim$~few Mpc, it is virtually
impossible for a proton of initial energy $\sim 10^{20}$~eV to travel distances greater
than $50-100$ Mpc without losing a large fraction of its energy~\cite{stecker}-\cite{yoshida}. 
This distance forms the boundary of the so-called ``GZK sphere''. As is well known, the
apparent paradox is that several cosmic ray experiments have detected a total of approximately
15 events apparently having $E > 10^{20}$~eV but with no currently obvious astrophysical source
within the local GZK sphere.

\noindent
Due to the small data sample presently
available, relatively little is known with certainty concerning these trans-GZK events.
In particular, unresolved are the fundamental questions 
concerning the composition/charge of the primaries
and how the distribution of arrival directions 
is related to the spatial distribution of sources.
Since there is no quantity analogous to galactic redshift to indicate relative
distances, statistical analysis of the arrival directions
of cosmic ray induced airshowers is necessarily
restricted to two dimensional tests over angular areas. And while the angular
distribution is obviously the projection of an underlying spatial distribution,
the danger of inferring spatial properties from angular characteristics is well known
due to the impossibility of deconvolving independent projection effects.

\noindent
What have been called ``small scale anisotropies'' as manifested by the apparently
nonrandom clustering of EHECR arrival directions in multiplets (doublets and triplets)
may or may not be correlated with 3-dimensional inhomogeneities.
Small scale angular anisotropies in the distribution of galaxies are associated with
localized spatial inhomogeneities, and this is similarly true for EHECRs if they
originate from sources embedded in a matter distribution (luminous or dark).
While it is reasonable to speculate this is the case, there is as yet no concrete 
evidence to support even this basic hypothesis. At this time, an equally viable hypothesis
is to associate the clustering with magnetic focusing in which case there is
no correlation with a 3-dimensional density distribution. It is also possible that
the observed distribution of arrival directions results from both (or other) causes.

\noindent
For this study, we analyze the AGASA data for airshowers with zenith angles
$\leq 45^\circ$ contained in references \cite{takeda} and 
\cite{hayashida} plus three additional events with $E > 10^{20}\,$eV listed on the 
AGASA website~\cite{ag_web}.
Inclusion of the three newest high energy events listed on the website is desirable
in order to increase the statistics of the tests used here, but is admittedly not 
optimal since it is currently not possible to include the additional events 
between $4\times10^{19} \;
\mathrm{and} \; 1\times10^{20}\,$eV that will be part of the next AGASA update. Where a bias
is possible due to this selective inclusion of events, the sample is restricted to only 
the 57 events contained in reference \cite{hayashida}.
Although it forms a doublet with a higher energy event and is close to the threshold cutoff
of $4\times10^{19}\,$eV, we specifically exclude the event with
$E = 3.89\times10^{19}$eV because it is impossible to assess the significance of this
additional pairing without having the positions of all other events of 
energies $3-4\times10^{19}\,$eV. 
In addition, although there are data for Haverah Park, Volcano Ranch, and Yakutsk events above 
$4\times10^{19}$eV, for the most part, we choose not to include them here since a large part
of our analysis involves energy cuts, and it remains uncertain as to how closely the energy 
estimates match between the various experiments. This limits the significance of the results,
but it is probably best to be conservative until, at the least, revised energies are published from
the re-analysis of the Haverah Park data~\cite{watson}.

\noindent
Previous investigations  of AGASA data have found statistically significant clustering on small 
scales with no apparent clustering or anisotropy on large scales; a representative sample of
model-independent studies are references \cite{takeda} and \cite{uchihori}-\cite{clay}.
For the present analysis, a robust angular
correlation estimator is applied to the AGASA EHECR arrival directions to probe for possible
departures from homogeneity on all angular scales.
The grouping of EHECR arrival directions into doublets and triplets
is re-analyzed with the Monte Carlo simulations used here yielding larger probabilities
for the presence of small scale clusters
but with similar significance compared to previously reported results.
A cross correlation estimator is then applied to energy-partitioned 
subsamples of EHECRs with the result that for the AGASA sample there appear to
be three distinct (uncorrelated) distributions with one being preferentially
located near the supergalactic equatorial plane. 

\section{The Two Point Angular Correlation Function}
\noindent
The two point correlation function is an effective and commonly used statistic 
to measure departures from homogeneity of an observed distribution of points. 
With no distance information available for cosmic ray events, the starting point is
the two point angular correlation function
representing the projection of the spatial correlation function
on the celestial sphere. It is defined in terms of the joint probability $\delta P$ for
finding two points separated by an angle $\theta$,
\begin{equation}
\delta P = n^2\left[ 1 + w(\theta)\right]\delta\Omega_1\delta\Omega_2 \; \; \mbox{,}
\label{Eq1}
\end{equation}
where $\delta\Omega_1$ and $\delta\Omega_2$ are solid angle elements containing
two points separated by angle $\theta$ and $n$ is the mean surface density of points;
$w(\theta) = 0$ for a homogeneous (uncorrelated) distribution of points
(for a detailed development of correlation functions see~\cite{peebles}). 
As a practical matter, $w(\theta)$ is computed from the ratio of counts of pair
separations for the data sample, \textit{DD}, and the average pair separations  
from a Monte Carlo generated test model, \textit{RR}, 
\begin{equation}
w(\theta) = \frac{DD}{RR} - 1 \; \; \mbox{.}
\label{Eq2}
\end{equation}
Previous investigations utilizing this estimator for analyzing cosmic ray arrival directions
include references \cite{tinyakov} and \cite{clay}.
\noindent
However, as shown by Landy and Szalay, the estimator in (\ref{Eq2}) generally exhibits variances
larger than the values expected when the pair counts follow a Poisson distribution~\cite{landy}. 
They analytically developed a modification 
that possesses a variance naturally free of a higher order term present in the
variance of the estimator in (\ref{Eq2}). In addition, the Landy-Szalay (LS) form for $w(\theta)$ is
robust in reducing spurious correlations resulting from inaccuracies in the assumed test model. 
The LS form for $w(\theta)$ is
\begin{equation}
w(\theta) = \frac{DD}{RR} - 2\frac{DR}{RR} + 1 \; \; \mbox{,}
\label{Eq3}
\end{equation}
where it can be seen that the data/random cross-term in (\ref{Eq3}) is responsible for the self-correcting
and stabilizing feature that reduces both the effects of modeling errors~(including boundary effects)
as well as the spurious correlations due to large data fluctuations intrinsic to small samples. 
When boundary effects are small, the data sample sufficiently large, and the test model
faithfully reproduces the actual average background distribution, the results of using equation
(\ref{Eq2}) are accurate and agree with (\ref{Eq3}). If the test model is approximately correct,
highly significant features will appear the same regardless of whether (\ref{Eq2}) or (\ref{Eq3}) 
is utilized. Here it is safest to use the LS form for $w(\theta)$
since the data sample of AGASA EHECRs is still relatively small and the
mean density of the underlying source distribution is unknown.

\noindent
The Monte Carlo sampled distribution of EHECR arrival directions 
is constructed to be uniform in right ascension and
follows the observed declination distribution 
for AGASA events with zenith angles $\leq 45^\circ$ 
as presented in~\cite{takeda}. The test model therefore corresponds
to a both isotropic and homogeneous source distribution convolved with the 
sky coverage and detector acceptance of the AGASA experiment. 
Detailed examination shows that the Monte Carlo random catalogs
accurately reproduce the polynomial fit to the declination distribution over the
entire interval of $-10^\circ \leq \delta \leq 80^\circ$.

\begin{figure}[t!]
\centering
\epsfig{figure=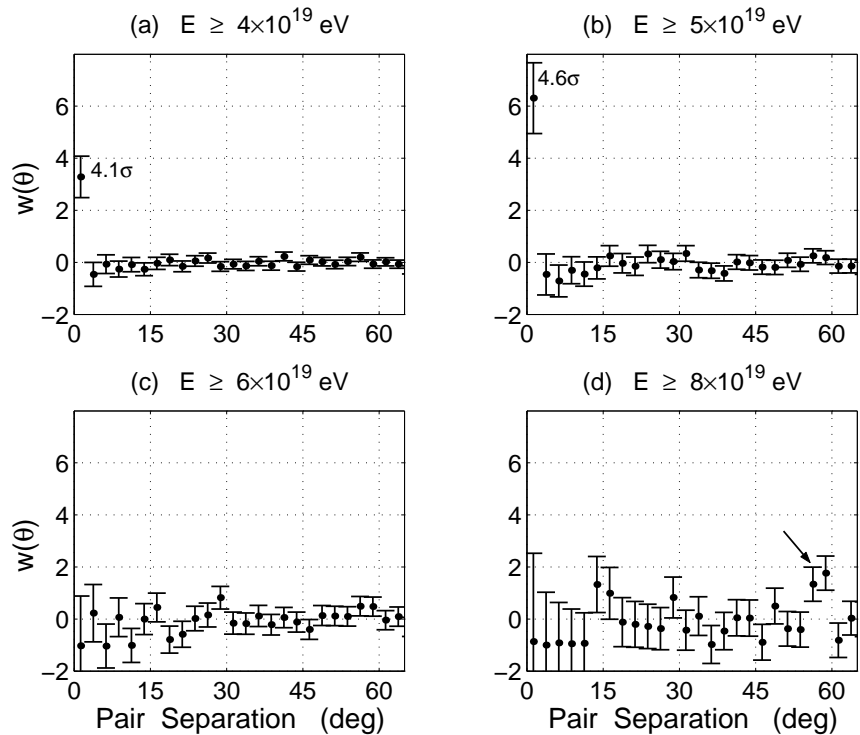,width=0.7\textwidth}
\caption{The two point correlation function for a sample of 60 AGASA events subject to energy cuts
for a bin size of 2.5 degrees. The two bins marked by the arrow in (d) combine to form a $3\sigma$
point for a bin size of $3^\circ$.}
\label{figure1}
\end{figure}

\noindent
The two point angular correlation function is shown Figure \ref{figure1}
for a bin size of $2.5^\circ$ and for various
energy cuts. The small angle clustering in Figures \ref{figure1}(a,b) is apparent from
the $> 4\sigma$ positive correlation in the first bin (with MC probability
of $\sim 0.1\%$). 
As shown, the plots extend only
to pair separations of $65^\circ$, but there are no significant correlations at larger separations
for any of the samples. 
For a bin width of $3^\circ$, the values of $w(\theta)$ for the first bin are slightly
reduced but with similar significance to those with the $2.5^\circ$ bin size. 
The positive correlation in the first bin drops very rapidly and disappears when the bin
width reaches $4^\circ \; \mathrm{to} \; 5^\circ$. The null correlations for bin sizes of 
$4^\circ \; \mathrm{and} \; 5^\circ$
approximate the results obtained from perturbing the positions around their stated values with
a Gaussian error distribution.
The lack of any statistically significant structure at the larger angles demonstrates that
when the sample is considered in its entirety,
the AGASA EHECR distribution of arrival directions is statistically homogeneous on all
angular scales except for those $< \, 4^\circ$ .

\noindent 
A further comparison of Figures~\ref{figure1}(a,b) shows that while there 
are three additional pairs in the entire sample 
compared to the sample cut at $E \, > \, 5\times10^{19}\:$eV (7 versus 4), the
value of $w(\theta)$ and the
significance of the small scale clustering is somewhat greater in the latter sample due to the
smaller total number of events (35 versus 60). The change in significance is related to the
fact that, for example, the probability for having zero pairs in a random sample is less for the
larger sample ($\approx 20\%$ for N = 60 and $\approx 60\%$ for N = 35). 
This result serves as a reminder that while
the clustering seen in the AGASA dataset 
is statistically significant, it is quite possible that some of the pairing in the sample
of 60 events is due to chance projection.  
Also of interest is the lack of clustering of events with
$E \, > \, 6\times10^{19} \:$ eV for the AGASA sample analyzed here
(0 pairs with separations $< \, 4$ degrees, 1 pair with separation $< \, 5$ degrees). 
However, with only 25 
events in this subsample, it is not possible to form any conclusions relative to this result. 

\noindent
The two bins highlighted by the arrow in Figure \ref{figure1}(d) combine to form
a $3\sigma$ result in a $3^\circ$ bin at $\sim 60^\circ$ for events with $E \, > \,
8\times10^{19}$ eV. 
As discussed previously, one of the strengths of the LS estimator
is that similar features appearing in results obtained from using the estimator in (\ref{Eq2})
are naturally suppressed by the cross term in (\ref{Eq3}). We have checked that this feature is 
not the result of
an average density anisotropy between different parts of the sky where the highest energy events
are located, but is due to the formation on the celestial sphere of two essentially equilateral
triangles with event arrival directions located at the vertices. Shifting the bin center
or increasing the bin size to $4^\circ$ strongly suppresses the correlation. Inclusion of the three newest 
AGASA high energy events has no effect on the significance of the point, but removal of the three events  
between $9 - 10\times10^{19} \,$eV reduces the correlation to insignificance. While the Monte Carlo
probability for such an occurrence is only $\sim 1\%$, it is suspected that this feature
is just an unusual chance alignment in the data. Thus, although it seems unlikely this
is a true correlation indicating structure, it is worth noting for further 
study as larger event samples become available.  

\section{Multiplet Counting and Probabilities}
\noindent
Evaluating the true significance of the observed clustering requires
additional consideration of the probability of occurrence.  
The two point correlation function involves pair counts regardless of whether
the pairs occur in doublets, triplets, quadruplets, etc. However, since the probability of 
random occurrence for three doublets can be very different than for one triplet, it
is important to analytically or numerically determine probabilities for various multiplet
configurations. This implies that the first bin of the autocorrelation function may or may not represent
the actual probability of a given clustering configuration. This is an expected result since the
two point correlation function is only the lowest order term in a Taylor series expansion
of the characteristic function of the probability distribution. The presence of 
triplets and higher order multiplets requires higher order correlation functions
to accurately estimate their probability of occurrence; however, the two point 
function is sufficient for exposing departures from homogeneity.

\noindent
Specific values of clustering probabilities depend on the definition
of a multiplet.
In the existing data from AGASA, Haverah Park, Yakutsk, and Volcano Ranch,
there are two triplets, each having all pair separations less than $2.5^\circ$.
Thus, in computing the probabilities presented here, a multiplet is defined as
a set of points within
a covering circle of diameter $\Delta\theta$ equal to the experimental 
uncertainty in angular position. 
This implies that the error circle of radius $\Delta\theta$ of each point in the multiplet
intersects the estimated positions of all the other points, or, equivalently, that
all $n(n-1)/2$ distinct pairs are separated by $\leq \Delta\theta$. 

\begin{figure}[t]
\centering
\epsfig{figure=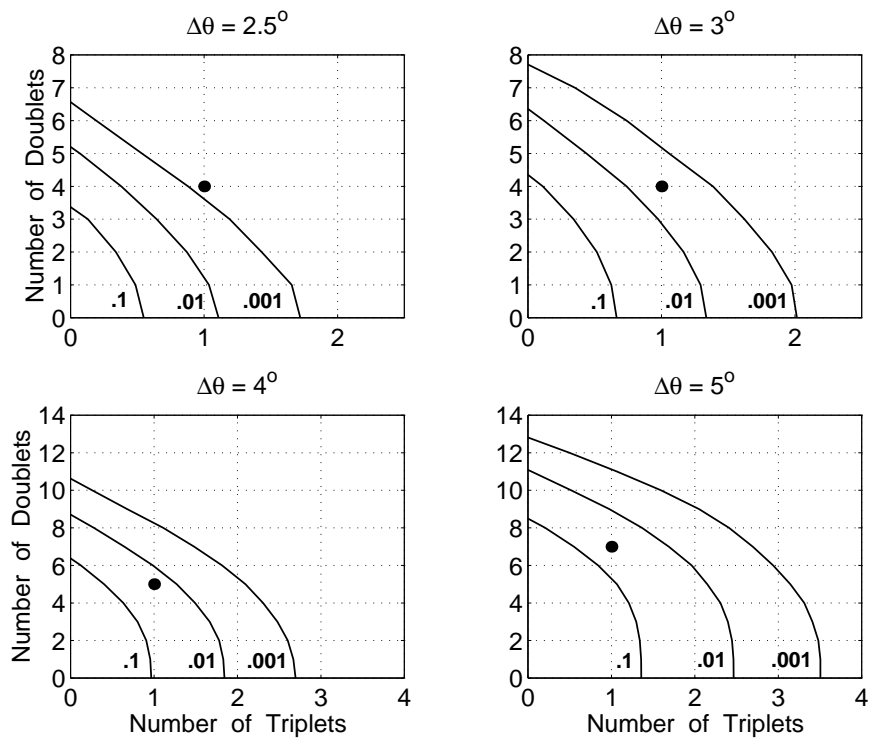, width=0.7\textwidth}
\caption{Joint probability contours for a sample size of 57 events for various opening angles.
The solid circles are the multiplet configurations of the corresponding AGASA dataset.
The probabilities are logarithmically distributed between contours.}
\label{jp_57}
\end{figure}

\noindent
Based on probabilities computed from simulation, the approach used here
to estimate the significance of small scale clustering into multiplets is similar,
but not identical, to that employed by others (e.g., see~\cite{uchihori}).
In both the data and the Monte Carlo catalogs, only \textit{distinct} multiplets are used
in the final results. That is, our counting algorithm first identifies all
multiplets from doublets through sextuplets for a specified opening angle (pair separation), 
and then correction factors are applied to each count to remove the contributions of higher 
order multiplets on lower orders.
This is straightforward, but sometimes unexpectedly subtle as, for example,
when four points combine to form two distinct triplets instead of one quadruplet
requiring five instead of six doublet pairs to be subtracted from the raw doublet count.
Depending on the sample size simulated, between 100,000 and 500,000 Monte Carlo trials
are used to obtain numerically converged probabilities. 
The random catalogs are constructed in the same way as for the correlation functions.
Details of our Monte Carlo sampling routines together with
a comparison of our results with \cite{uchihori} are contained in reference \cite{omalley}.

\begin{figure}[t]
\centering
\epsfig{figure=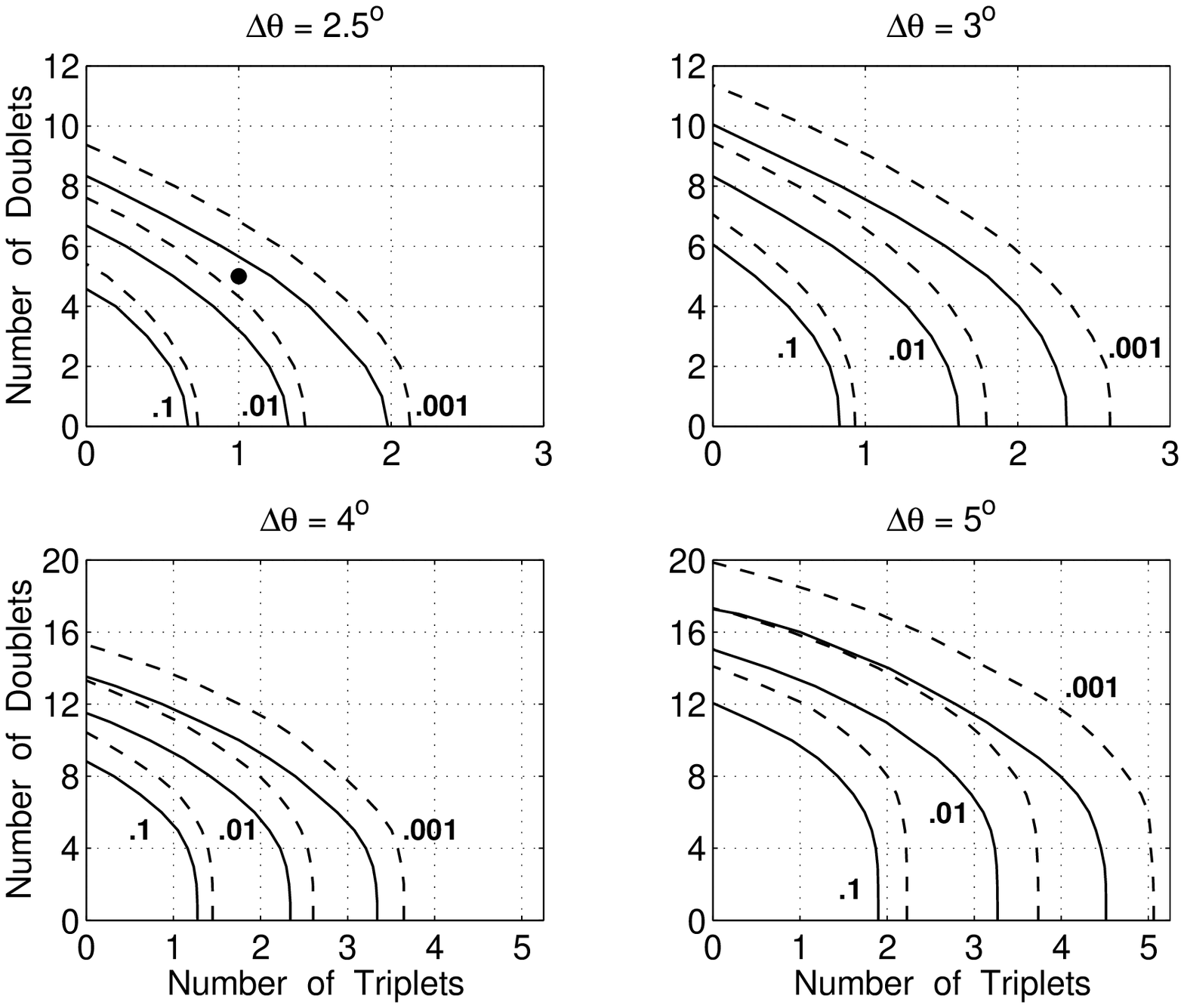, width=0.7\textwidth}
\caption{Joint probability contours for sample sizes of 72 (solid) and 80 (dashed) 
with the solid circle being the multiplet configuration of the corresponding 72 event 
AGASA dataset. The probabilities are logarithmically distributed between contours.}
\label{jp_72_80}
\end{figure}

\noindent
Analytical estimates of clustering probabilities based on Poisson statistics have
been presented previously in, e.g.,
Dubovsky \textit{et al.}~\cite{dubovsky}. Goldberg and Weiler 
constructed analytical formulas for exact probabilities and asymptotic forms 
that differ naturally from the pure Poisson
estimates due to the presence of system constraints~\cite{goldberg}.
Further results from using the Goldberg-Weiler formulation are contained in 
Anchordoqui \textit{et al.}~\cite{anchordoqui}.
The method is based on calculating
the probabilities of bin occupation numbers for specified
bin widths and numbers of bins, and yields probabilities for distinct occupation
configurations under the assumption that all bins have equal occupation probabilities 
(an approximation to the non-uniform acceptance of ground array detectors such as
AGASA). The bin area is chosen to be $\pi\Delta\theta^2$ for a correlation (opening) angle
of $\Delta\theta$ (leading to a bin diameter of $2\Delta\theta$). 

\begin{figure}[t]
\centering
\epsfig{figure=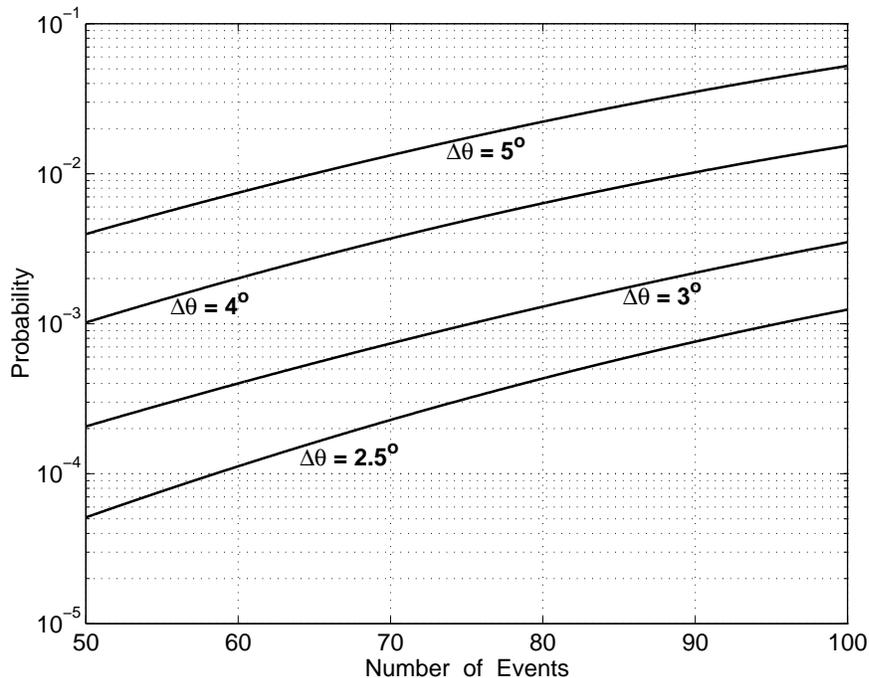, width=0.7\textwidth}
\caption{Inclusive quadruplet probabilities $P(quadruplet\geq 1)$ for the AGASA declination
distribution as a function of event sample size for various opening angles.} 
\label{quad_prob}
\end{figure}

\noindent
For exclusive doublet configurations with zero higher order multiplets, the pair separation
definition of a multiplet used in this study is equivalent to the bin occupation probability
given by the Goldberg-Weiler formula.
For a simulated uniform declination
distribution, the exclusive doublet probabilities differed by less than 10\% from the
analytically computed values and were within 30-40\% when using the AGASA declination 
distribution. Exclusive triplet probabilities from simulation were also consistent with
the analytical results as were the joint exclusive doublet/triplet probabilities although
differences begin to appear more significant when compared with the non-uniform AGASA
declination distribution. 
In general, with an appropriate choice for the total solid angle used in the
analytical formula, the differences between the two
approaches can be less than a factor of two for all doublet/triplet 
configurations. 
As expected, by requiring all distinct pair separations to be $\leq \Delta\theta$, 
the quadruplet probabilities shown in Figure \ref{quad_prob} 
are smaller by factors of 5-10
compared to those computed analytically where the bin diameter is $2\Delta\theta$. 

\noindent
Joint probability contours have been constructed 
for various doublet/triplet combinations as a function of sample size and opening angle
from the simulations utilizing the AGASA declination distribution. 
For the sample sizes considered here, the probabilities for quintuplets,
sextuplets, etc. are negligible ($P \, \lesssim \, 10^{-5}$) compared to the doublet and triplet
probabilities
with the quadruplet probabilities also being safely neglected ($P \, \lesssim \,  10^{-3}$) for
$\mathcal{N}_{sample} \;  \lesssim \; 75$ when the opening angle is $\lesssim \, 3$ degrees.
Examples are shown in Figures \ref{jp_57} and \ref{jp_72_80} for 
contours representing the joint inclusive probabilities 
\begin{equation}
P(N_{doub} \geq N^{0}_{doub}) \: \cap \: P(N_{trip} \geq N^{0}_{trip})
\label{Eq4}
\end{equation}
given specified values of $N^{0}_{doub}$ and $N^{0}_{trip}$.

\noindent
Consideration of the contours in Figure \ref{jp_57} for a bin size of $2.5^\circ$ shows
that the observed configuration of 4 doublets and 1 triplet for an opening angle of $\Delta\theta \:
\leq \: 2.5^\circ$ has a probability of less than 0.1\% to occur due to random projecton.
Further examination reveals
common qualitative characteristics of clustering probabilities.
For example, note that although the probability for finding $\geq 2$ triplets and $\geq 0$ doublets
is much smaller than for $\geq 6$ doublets and $\geq 0$ triplets, there are cases where a triplet/doublet
combination has the same or even greater probability as the same number of pairs distributed only
in doublets. 
It can also be seen in this figure that just as in the correlation function,
the significance of the probability rapidly decreases as the opening angle increases.
The contours for the larger sample sizes in Figure \ref{jp_72_80} 
may be used to estimate the statistical
significance of doublet/triplet clustering in future AGASA event samples.
The inclusive probabilities for quadruplets as a function of sample size $N \leq 100$ 
for opening angles of $2.5^\circ$, $3^\circ$, $4^\circ$, and $5^\circ$ are shown 
in Figure \ref{quad_prob}. 

\noindent
We note that the probability distributions obtained for doublets
and triplets for a combination of AGASA, Haverah Park, and Yakutsk declination 
distributions are similar in shape to those obtained in reference \cite{uchihori}.
However, a direct comparison of the probabilities found here shows them to be a factor of 2 larger
(twice as likely) for the total number of uncorrected doublets (i.e., total number of pairs without  
regard to what order of multiplet they appear in), 
and a factor of 1.5 larger for uncorrected triplets for the 92 events considered. 
The results of Uchihori \textit{et al.} indicated that the small angle clustering for the entire sample
of 92 events was only marginally significant for an opening angle of $3^\circ$ and not significant
for larger opening angles~\cite{uchihori}. 

\begin{table}[t]
\caption{Summary of joint inclusive probabilities for distinct multiplet configurations of EHECR arrival
directions from AGASA, Haverah Park, Yakutsk, and Volcano Ranch. The table entries
referring to ``low energy'' are for events with $E < 5\times10^{19} \,$eV.}
\label{table1}
\begin{center}
\begin{tabular}{|c|c|c|c|c|} \hline
Sample & $\hspace{1em} \Delta\theta$ \hspace{1em} & Doublets 
& Triplets & MC Probability \\ \hline\hline
N = 102, all events \hfill & 3 & 8 & 2 & 1.0\% \\ \hline
\hfill & 4 & 11 & 2 & 8.6\%\\ \hline\hline
N = 91, exclude HP low energy \hfill & 3 & 5 & 2 & 0.7\%\\ \hline
\hfill & 4 & 7 & 2 & 6.5\%\\ \hline\hline
N = 56, exclude all low energy \hfill & 3 & 0 & 2 & 0.1\%\\ \hline
\hfill & 4 & 1 & 2 & 0.6\%\\ \hline
\end{tabular}
\end{center}
\end{table} 

\noindent
Although our probabilities for total (uncorrected) pair counts 
are greater, when considering the corrected joint doublet/triplet probabilities for distinct
multiplets only, we obtain Monte Carlo probabilities numerically
similar to those in \cite{uchihori}. Values of the same order are obtained for the extended
sample of 102 events with the results summarized in Table \ref{table1}. From this table it
can be seen that as in the sample of AGASA events only, the amount of clusters decreases
but the significance of the clustering increases as lower energy events are removed from
consideration. For example, for $\Delta\theta = 4^\circ$, 
ignoring the likely differences in energy scales between the experiments and arbitrarily
removing all events with reported 
energies $< 5\times10^{19} \,$eV eliminates
all but 1 doublet but leaves the 2 triplets with a resulting MC probability of $0.6\%$. 
Thus, it is possible that the small angle clustering is more significant in the
combined sample than previously thought; however, it must be emphasized that this cannot be confirmed
until the revised energies are given for the Haverah Park events.

\begin{figure}[b!]
\centering
\epsfig{figure=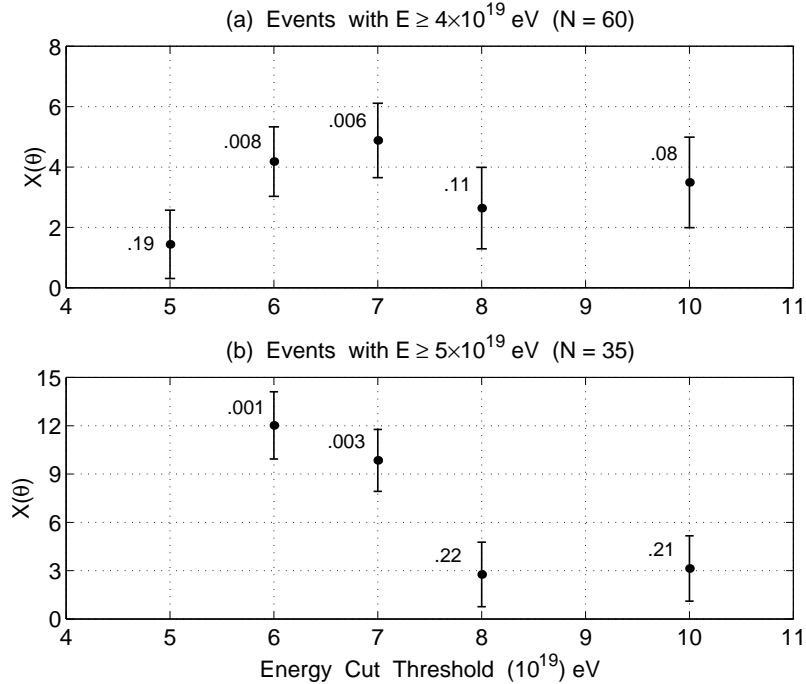,width=0.65\textwidth}
\caption{Cross correlation function values in the first bin for the AGASA
events above and below various energy thresholds (bin size = $2.5^\circ$). 
The numbers indicate the MC probability of occurrence where highly correlated
samples have small probabilities and vice versa.}
\label{ag60_xcut1}
\end{figure}

\section{Cross Correlations of Energy Partitions}
\noindent
The two point correlation function in Section 2 is the autocorrelation function for
a sample of data points. It is often of interest to compute the cross correlation
of two data samples. In this case, the basic form analogous to (\ref{Eq2}) is
\begin{equation}
\mathcal{X}(\theta) = \frac{D_1D_2}{R_1R_2} - 1 \; \; \mbox{,}
\label{Eq5}
\end{equation}
where the subscripts 1 and 2 refer to the different samples. It is straightforward to construct
an LS estimator analogous to ({\ref{Eq3}) that reduces to (\ref{Eq3}) when $D_1 = D_2$, 
\begin{equation}
\mathcal{X}(\theta) = \frac{D_1D_2}{R_1R_2} - \frac{1}{2}\frac{(D_1R_1 + D_1R_2 + D_2R_1 + D_2R_2)}{R_1R_2} 
+ 1 
\; \; \mbox{,}
\label{Eq6}
\end{equation}
and where $\mathcal{X}(\theta)=0$ implies the two samples are statistically uncorrelated. 

\noindent
As a starting point, the values for the first bin of $\mathcal{X}(\theta)$ 
for a bin size of $2.5^\circ$ are shown in
Figure \ref{ag60_xcut1}(a) above and below a given energy threshold for the 60 AGASA 
events. As there are no events in this dataset
with energies between $8 \, \textrm{and} \, 9\times10^{19} \,$eV, the results for
the energy cut at $8\times10^{19} \,$eV are the same as for $9\times10^{19} \,$eV.
Although the first bin values of
$\mathcal{X}(\theta)$ are positive for all cases, only those corresponding to
$< 1$\% probability are considered significant. Thus,
the data indicates that events below and above $5\times10^{19} \,$eV are
not strongly correlated, and the same holds true for the threshold energy $8\times10^{19} \,$eV.
As shown in Figure \ref{ag60_xcut1}(b),
after removing the 25 events with $E < 5\times10^{19} \,$eV, the independence
of the events below and above $8\times10^{19} \,$eV and the strong autocorrelation of the
events with $5 \leq E < 8\times10^{19} \,$eV are even more apparent.
These results suggest the division into three groups with energies
$4 \leq E < 5\times10^{19} \,$eV, $5 \leq E < 8\times10^{19} \,$eV, and
$E \geq 8\times10^{19} \,$eV.

\begin{figure}[t]
\centering
\epsfig{figure=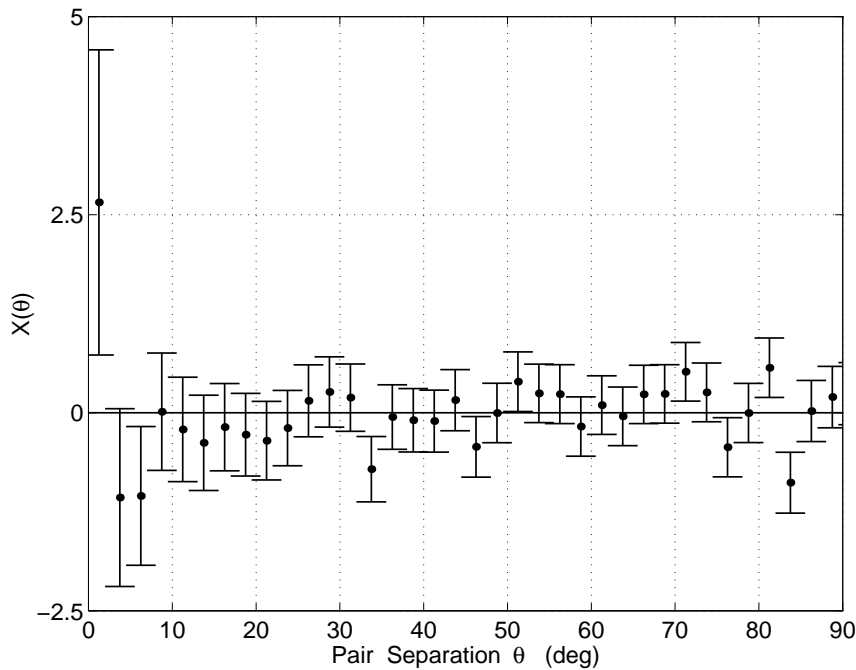,width=0.7\textwidth}
\caption{The cross correlation function for AGASA events with $5\times10^{19} \leq  
E \leq 8\times10^{19} \; \mathrm{and} E > 8\times10^{19}\,$eV 
showing that these two distributions are largely uncoupled at all angles
(the $1.3\sigma$ positive correlation value in the first bin is
statistically insignificant).}
\label{xcor5_8}
\end{figure}

\noindent
Using the three energy partitions as defined above, the cross-correlation functions are computed
for the three possible combinations. Figure \ref{xcor5_8} shows the statistical independence
at all angles of the middle and high energy partitions. The results for the
other two cases are very similar with the exception of two $3\sigma$ points appearing at 
angular separations of $\approx 45^\circ \; \textrm{and} \; 70^\circ$ in the cross-correlation of
the $E < 5\times10^{19} \,$eV and the $E \geq 8\times10^{19} \,$eV partitions. For both points, the
correlation values in the immediately adjacent bins are negative for the $2.5^\circ$ bin width.
This suggests that both points are probably unusually
large fluctuations similar to the large scale feature in the 2-point function described in Section 2.
As expected for this type of feature, the values for these points using (\ref{Eq6}) are reduced 
compared to those obtained from using (\ref{Eq5}). 

\noindent
Of course, more data is required to confirm the existence and
independence of these energy partitions. If confirmed, the implications are 
that events separated in energy are being influenced differently or are of different
origin or composition. This could be especially important in helping to determine the nature
of the trans-GZK distribution above $8 \times10^{19} \,$eV. 

\section{Anisotropy and the Supergalactic Plane}
\noindent
As noted, the correlation function reveals no compelling departures from homogeneity on large angular
scales for either the entire or the energy-partitioned samples. It is then natural to check whether
the events appear to show directional dependences with respect to coordinate
systems such as galactic or supergalactic. Testing for anisotropy begins with sector counts
of data compared to expected values from simulation. This is
sufficient for the analysis here although it should be noted that it is possible 
to describe deviations from isotropy by constructing
statistical estimators such as plane enhancement factors. 

\begin{figure}[t]
\centering
\epsfig{figure=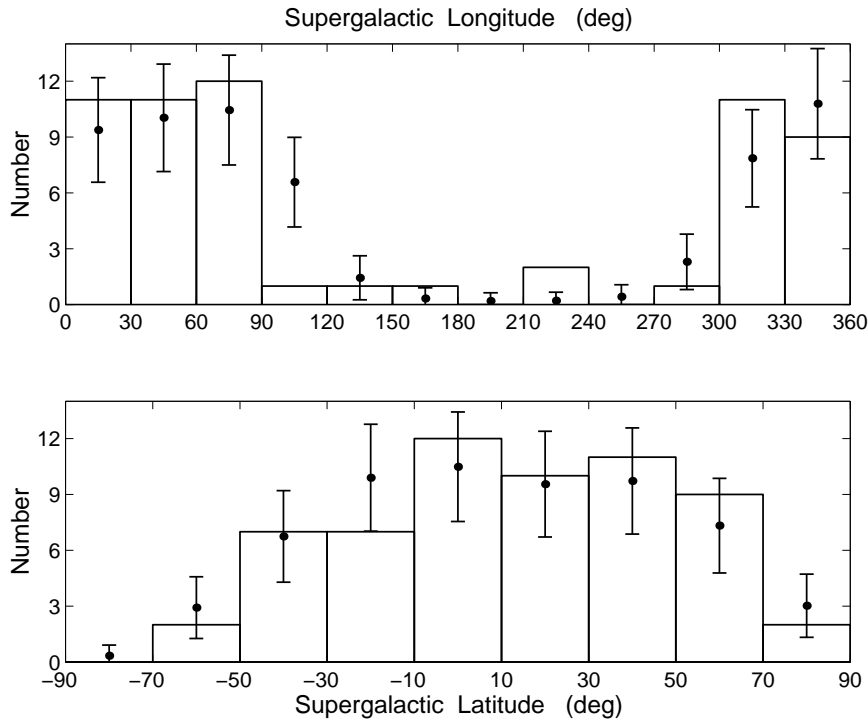,width=0.7\textwidth}
\caption{The distribution of 60 AGASA events in supergalactic coordinates. The
points marked by solid circles are the expected values obtained from the Monte Carlo
simulation under the assumption of isotropically and uniformly distributed
sources. The 1-sigma error bars are derived from the simulation results and are approximately
Poisson.}
\label{figure7}
\end{figure}

\noindent
Figure \ref{figure7} shows the distribution in supergalactic (SG) coordinates  of the 60 AGASA 
events where it can be seen that the arrival directions
appear to be isotropically distributed. 
However, referring to Figure \ref{sgb_part}, constructing similar histograms
for the three partitions described in the previous Section reveals that the group with energy
$5 \, \leq E < \, 8\times10^{19} \,$eV is aligned 
with the SG equatorial plane at the $3\sigma$ level (corresponding to
a $0.6\% \;$ probability). The lower and higher energy groups
are consistent with statistically isotropic distributions although the $2.5\sigma$ deficit near the SG 
equator for the low energy group should be checked in future studies. The
SG longitude distributions for the three energy-partitioned samples cannot be distinguished from
isotropically distributed sources. 

\begin{figure}[t]
\centering
\epsfig{figure=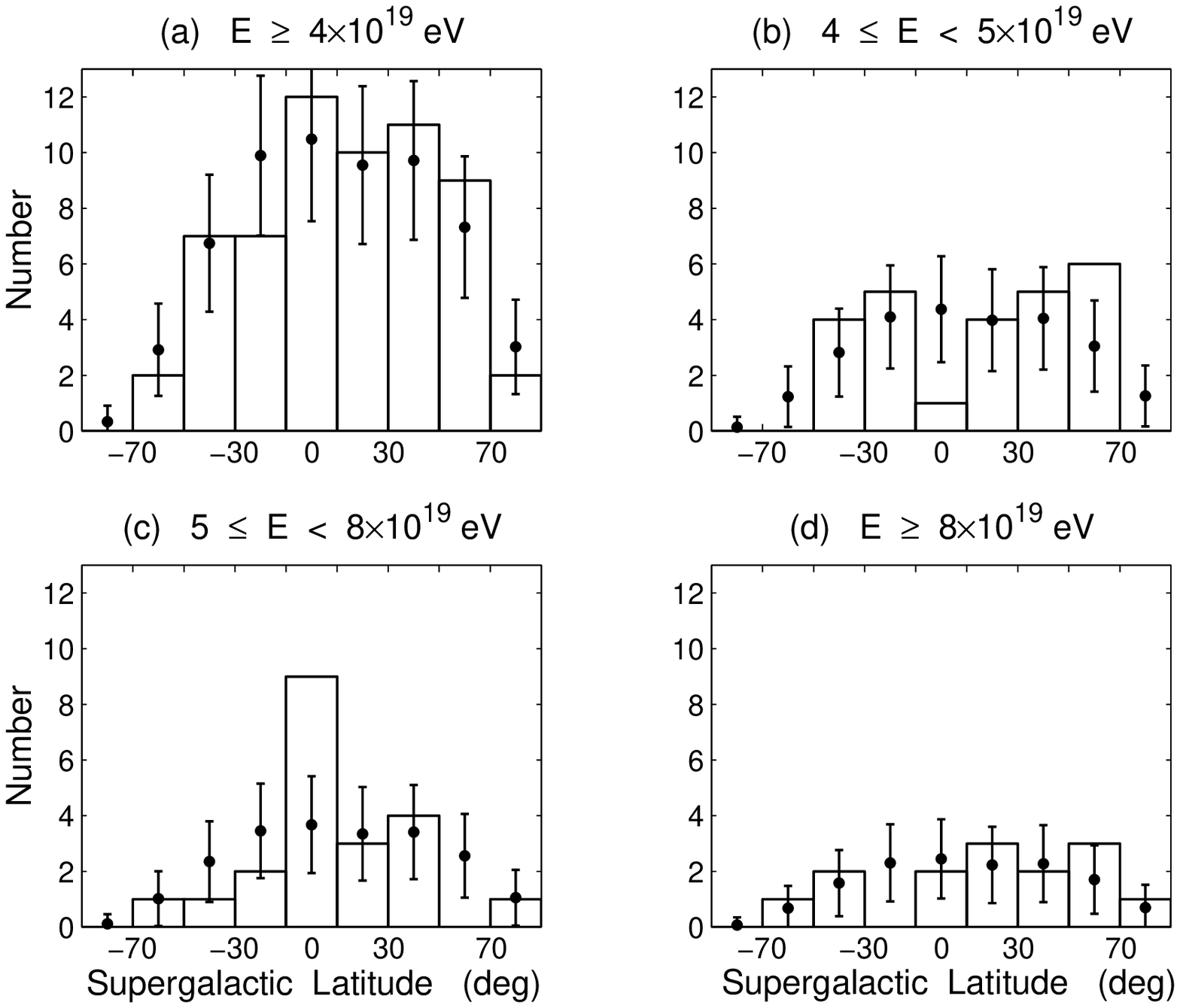,width=0.7\textwidth}
\caption{The supergalactic latitude distribution of 60 AGASA events partitioned by energy. 
The points marked by solid circles are the expected values obtained from the Monte Carlo
simulation under the assumption of isotropically and uniformly distributed sources.
The excess of events in (c) from $-10^\circ \leq SGB \leq 10^\circ$ is at the $3\sigma$ level 
(9 observed versus $3.68 \pm 1.74$ expected) with a Monte Carlo 
probability of 0.6\%.}
\label{sgb_part}
\end{figure}

\noindent
As an independent test of the significance of the alignment of the $5 \leq E < 8\times10^{19} \,$eV
group with the SG plane, the energies of the 60 AGASA events were randomly shuffled among the
(fixed) event coordinates. Aftr 100,000 trials, there were an average of $4.2 \pm 1.5$ events with
$-10 \leq SGB \leq 10^\circ$, and in only 0.2 - 0.3\% of the trials did the number equal or exceed the
observed value of 9.
This probability is of the same order as the 0.6\% found from MC sampling
the coordinates, and confirms the statistically significant nonrandom alignment of this partition
with the SG equatorial plane.

\noindent
A possible correlation between EHECRs and the Local Supercluster 
was hypothesized by Stecker in 1968 based on his calculations of propagation paths 
attenuated by photomeson production~\cite{stecker}.
Stanev \textit{et al.} noted in 1995 that EHECR arrival directions show a correlation with the
SG plane, and that this was consistent with the hypothesis that powerful radio sources within the
Local Supercluster, being concentrated in this plane, could be the production sites of high energy
cosmic rays~\cite{stanev}. Following this, other studies have pointed out that some of the
multiplets are preferentially aligned in this direction~\cite{takeda,uchihori,hayashida_2}. 
Because the AGASA triplet is
within a few degrees of the SG equator, and the energies of the triplet are
between $5 \; \textrm{and} \; 8\times10^{19} \,$eV, the statistical significance of the results
are strongly influenced by this structure. Due to the anticipated relative differences in
energy scales, it is not
possible at this time to further pursue the correlation of the arrival directions with the SG
equatorial plane as a function of energy utilizing the 
combined data from other experiments.

\noindent
If the observed anisotropy and correlation with the SG plane of the middle energy group is
in fact related to a source distribution residing within the Local Supercluster, then it is
probable that this same energy group should also be anisotropically distributed with respect
to SG longitude. Specifically, with our Galaxy being located toward the periphery of the Local
Supercluster, there are many more galaxies along the Galaxy line-of-sight direction corresponding
to $SGL \approx 0^\circ$ (toward the Virgo cluster which lies at the center of the Local Supercluster)
than along the opposite direction of $SGL \approx 180^\circ$. 
However, inspection of Figure 7 reveals that there is no AGASA coverage along SGL values
where deficits would be expected to exist.
Thus, determination of this possible anisotropy requires
data from a different experiment and cannot be analyzed from the current dataset.

\section{Summary}
\noindent
In the above analysis, we consider features with $\gtrsim \, 3\sigma$ values 
($\lesssim \, 1\%$ MC probability) to be potentially indicative of structure 
or to warrant further examination with larger datasets. Features at the $\gtrsim
\, 2-2.5\sigma$ level also merit further attention, but it is premature to
attribute significance to them in this study.
The results may be summarized as follows:\\
\indent 1. The AGASA EHECR arrival directions
continue to exhibit clustering on scales $\lesssim \, 3^\circ$ at the $4\sigma$ level 
with probabilities of random occurrence of $\sim .1 - .5$\%, and
continue to show no compelling departures from homogeneity on
scales $\gtrsim \, 5^\circ$,\\
\indent 2. Partitioning the AGASA sample by energy yields three uncorrelated
groups with $E < 5\times10^{19} \,$eV, $5 \leq E < 8\times10^{19} \,$eV, and
$E \geq (8-10)\times10^{19} \,$eV,\\
\indent 3. The partition with $5 \leq E < 8\times10^{19} \,$eV appears to be preferentially
aligned with the SG plane with a probability for random occurrence of $< 1\%$ 
while the other two partitions are statistically
consistent with isotropic distributions. Supporting evidence that this anisotropy is related
to the Local Supercluster would be an observation of an anisotropy in the numbers
of events toward SG longitudes in the neighborhoods of 
$0^\circ \: \mathrm{and} \: 180^\circ$, but this is not yet
possible due to the limited sky coverage of the AGASA experiment,\\
\indent 4. Combining the 102 events reported to have energies $> 4\times10^{19} \,$eV from AGASA,
Haverah Park, Yakutsk, and Volcano Ranch and considering
inclusive joint probabilities of distinct multiplet configurations, we find MC probabilities of
$\lesssim 1\%$ and $5-10\%$ for opening angles of $3^\circ$ and $4^\circ$, respectively.\\  
\indent 5. Without regard to differences in energy scales between the four experiments, 
eliminating all events in the 102 event sample with energies reportedly $< 5\times10^{19} \,$eV
yields a clustering probability of $0.1\%$ for the (0 doublet, 2 triplet) configuration observed
for an opening angle of $3^\circ$ and a $0.6\%$ probability for the (1 doublet, 2 triplet)
configuration observed for an opening angle of $4^\circ$.\\
\noindent
The change in significance of the clustering probability for the 102 event sample in making
the energy cut and reducing the observed configuration from 11 doublets and 2 triplets 
to 1 doublet and 2 triplets ($\Delta\theta = 4^\circ$) underscores the unlikely chance for
observing 2 triplets in a sample containing only $\sim 50-60$ events.

\noindent
It should also be noted that the values of the energy partitions obtained for the AGASA
events may or may
not correspond to an absolute energy scale as it is possible that the AGASA energy calibration
could be revised in the future. Nevertheless, the energies found here are consistent with
certain expectations of regimes where GZK or magnetic field effects change their relative importance.
And, of course, being based on relatively few events, these results can be significantly 
affected by the addition of the next 15 to 20 events.

\noindent
Beyond the characterization of statistical properties, a natural goal of this type of analysis
is to determine underlying causes. With so little known concerning the nature of the EHECR
distribution, it is tempting but difficult to extrapolate statistically significant features
of small datasets to physical properties of the parent distribution. Here that means 
attempting to associate the above results with repeating sources, changes in primary composition
or source distribution, or magnetic lensing. The difficulty is compounded by the lack of
knowledge of how many physical mechanisms are influencing the observed distribution. 
For example, it is plausible that if there exist neutral EHECR primaries unaffected
by GZK losses or magnetic fields, then there also exist charged primaries that are so
affected. In that event, a single type of production mechanism gives an observed spectrum that is
a superposition of charged primaries from sources residing within the local GZK sphere, and
neutral primaries that may originate from sources both within and beyond the GZK sphere. 

\noindent
What is fair to
hypothesize based on the results here is the existence of three distinct
energy distributions (presumably of extragalactic
origin) comprising the observed spectrum of EHECR events and that may possess the following
properties:\\
\indent 1. Cosmic rays with $E \; \lesssim \; 5 \times 10^{19} \; \mathrm{eV}$ comprising the high energy
tail of a primary distribution largely unaffected by GZK losses,\\
\indent 2. Primaries, possibly protons, with $5 \; \lesssim \; E \; \lesssim \;
8\times 10^{19} \,$eV that have lost energy through the GZK effect, and, with arrival
directions being aligned with the supergalactic plane, may originate
from sources located within the Local Supercluster,\\
\indent 3. trans-GZK primaries of unknown composition and origin with energies $E \; \gtrsim
\; 8\times10^{19} \; \mathrm{eV}$ that may or may not be losing energy via the GZK effect.\\
The consistency of this interpretation can be checked further once the energy scales of existing
datasets are brought into accordance, but confirmation or rejection probably requires a larger
single dataset as will be provided by Auger.

\vspace{5mm}
\large{\noindent \textbf{Acknowledgements}}\\*[0.5cm]
\small
The authors thank A.~Szalay for referring them to the use of the
Landy-Szalay two point correlation estimator, W.~Rindler for helpful suggestions, 
A.~Watson for information concerning Haverah Park data and for allowing access to
the high energy HP events, and 
Y.~Uchihori for his correspondence on some of the issues discussed in this paper
as well as help with data from Haverah Park, Volcano Ranch, and Yakutsk.
We would also like to thank H.~Goldberg and T.~Weiler for correspondence
concerning their analytical formula for computing clustering probabilities.
We are especially grateful to J.~Krizmanic for several informative discussions
together with his critical review of portions of this paper. Finally, we would
like to thank an anonymous referee for several helpful suggestions.

\end{document}